\documentstyle{article}
\newcommand {\vs}{\vspace{2mm}}
\newcommand {\hs}{\hspace}
\newcommand {\beq}{\begin{equation}}
\newcommand {\eeq}{\end{equation}}
\newcommand {\bit}{\begin{itemize}}
\newcommand {\eit}{\end{itemize}}

\newcommand {\cen}{\centerline}
\newcommand {\bcen}{\begin{center}}
\newcommand {\ecen}{\end{center}}

\begin{document}
\large
\setlength {\baselineskip}{0.cm}

\vspace*{2cm}
\cen{\bf The Molecular Front in Galaxies II :
Galactic-scale Gas Phase Transition of HI and H$_2$}
\vs

\cen{\bf Mareki {\sc Honma}, Yoshiaki {\sc Sofue} \& Nobuo {\sc Arimoto}}
\vspace{1cm}

\cen{ Institute of Astronomy, University of Tokyo, Mitaka, Tokyo 181, Japan}
\cen{ E-mail: honma@astron.s.u-tokyo.ac.jp}
\vspace{4.5cm}

\hs{2cm} {\bf Submitted to :} {Astronomy and Astrophysics Main Journal}
\vs

\hs{2cm} {\bf Running head :} {The Molecular Front in Galaxies II}
\vs

\hs{2cm} {\bf Section :} {Extragalactic astronomy}
\vs

\hs{2cm} {\bf Thesaurus :} {09.01.2; 09.13.02; 11.05.2; 11.09.4; 11.19.2}
\vs\vs

\hs{2cm} Proofs should be sent to M.Honma.

\newpage
\cen{\bf abstract}

We have examined the distribution of HI and H$_2$ gases in four face-on
galaxies by using the observed data of CO and HI line emissions from the
literatures.
We demonstrate that the gas phase transition of HI and H$_2$ occurs suddenly
within a narrow range of radius, which we call the molecular front.
We have tried to explain such phase transition in galactic scale with a help of
the phase transition theory proposed by Elmegreen.
The crucial parameters for determinating the molecular fraction $f_{\rm mol}$
are interstellar pressure $P$, UV radiation field $U$, and metallicity $Z$, and
we have constructed a model galaxy in which $P$, $U$ and $Z$ obey an
exponential function of the galacto-centric radius.
The model shows that the molecular front must be a fundamental feature of
galaxies which has an exponential disk, and that the metallicity gradient is
most crucial for the formation of the molecular front.
We have also tried to reproduce the observed molecular fraction $f_{\rm mol}$
by giving the set of ($P$, $U$, $Z$) observationally, and show that the model
can describe the variation of the molecular fraction $f_{\rm mol}$ in galaxies
quite well.
We discuss the implication of the molecular front for the chemical evolution of
galaxies.

\vspace{2cm}

{\bf keywords:} galaxies  - atomic gas - molecular gas - gas phase transition -
evolution of galaxies

\newpage
\section{Introduction}

The HI and H$_2$ gases are the main components of the interstellar matter (ISM)
and both are tightly related to the star forming activity in galaxies.
Therefore, investigating these neutral components of the ISM is one of the most
powerful ways to understand the structure and evolution of galaxies.
The HI gas can be observed directly in the HI-21cm line emission and H$_2$ can
be investigated by observing the CO-2.6mm line, while it is difficult to
observe H$_2$ molecules directly.
So far we have observed many galaxies in the CO-2.6mm line emission using the
Nobeyama 45m telescope (e.g. Sofue \& Nakai 1993; 1994; Sofue 1994).
Many other researchers have observed galaxies in the HI-21cm line using VLA and
other facilities (e.g. Rupen 1991; Irwin \& Seaquist 1991).
However, in spite of a large number of observations of galaxies in CO and HI,
there has been yet no direct comparison of the distribution of HI and H$_2$
with sufficiently high spatial resolutions.

We have made the first detailed comparison of the distributions of HI and
H$_2$(calculated from CO) in several edge-on galaxies (Sofue, Honma \& Arimoto
1994, hereafter Paper I) based on derivations of the density distributions of
HI and H$_2$.
We have obtained the radial variation of the molecular fraction ($f_{\rm mol}$)
which is defined by the ratio of molecular gas density to total hydrogen gas
density as,
\beq
f_{\rm mol}=\frac{2\times n({\rm H_2})}{n({\rm H})+2\times n({\rm H_2)}}.
\eeq
By making such quantitative comparison, we have demonstrated
1) that the central region within a few kpc is totally molecular,
2) that the outer region is totally atomic,
and 3) that the HI-H$_2$ gas phase transition occurs rather sharply at a
certain radius, which we shall call the molecular front.
The disk of a galaxy is divided into two radial zones at the molecular front;
the inner H$_2$ disk and the outer HI disk.

On the other hand, a theoretical study on the phase transition between HI and
H$_2$ has been made by Hollenbach et al.(1971) in rather small scale
corresponding to individual sizes of clouds and has recently been developed by
Elmegreen (1993) in galactic scale.
However, there has been no study yet that argues how $f_{\rm mol}$ varies with
the distance from the galactic center or whether a feature like the molecular
front can be theoretically reproduced.
In this paper we construct a gas phase transition model for disk galaxies and
investigate the variation of $f_{\rm mol}$ with the galacto-centric radius.
We show that the observed features mentioned above, including the molecular
front, can be reproduced by the model.

\section{The Molecular Front in Face-on Galaxies}

In Paper I, we obtained the distributions of HI and H$_2$ gases in four edge-on
galaxies by deconvolving position-velocity diagrams under the assumptions of a
flat rotation curve and an axisymmetric distribution.
In order to avoid the uncertainties which might come from these assumptions, we
now investigate the distributions of HI and H$_2$ gases in nearby face-on disk
galaxies and see whether the feature like the molecular front claimed in Paper
I really appears in face-on galaxies.
 From many face-on galaxies so have been observed in HI and CO, we select a
sample of four galaxies which are near enough to get sufficiently high
resolution and are investigated well to obtain other properties such as the
surface brightness and the metallicity gradient.
Our sample consists of four galaxies;
M51, M101, NGC 6946 and IC342,
which are typical late-type spirals classified as Sc, and many fundamental
observational data are available including HI, CO, optical luminosity profile
and oxygen abundance of HII regions.
The parameters for these galaxies and the references are summarized in Table 1.

\cen{Fig. 1.a - 1.d}

We show the surface density distributions of HI and H$_2$ gases in these four
galaxies in Fig.1.a - 1.d, where the amount of H$_2$ is calculated from CO
intensity using a constant conversion factor of 3$\times$10$^{20}$ (H$_2$ (K
km/s)$^{-1}$ ).
We conventionally assume that the conversion factor is universal and is the
same as the Galactic value.
The gas distribution for M101 has been azimuthally averaged, while those for
other galaxies have been obtained by averaging the distributions along the
major axis in both sides.
All galaxies in Fig.1 show conspicuous concentration of molecular gas at the
galactic center with the surface mass density of more than 100 M$\odot$/pc$^2$.
The density of molecular gas decreases steeply with the radius, showing local
enhancement in the arms in the cases of M51 and IC 342.
On the other hand, the distribution of HI gas is rather flat, almost constant
through the galactic disk.
The enhancement of HI density in the arm is not so clear as seen in the H$_2$.
This implies that the gas might be converted into H$_2$ once the gas density
becomes larger than the critical density which is around 5 M$_\odot$/pc$^2$.

\cen{Fig. 2.a - 2.b}

In Fig.2.a we show the molecular fraction $f_{\rm mol}$ which is defined by
equation (1).
We can see that $f_{\rm mol}$ at the galactic center is almost unity in any of
these galaxies and that $f_{\rm mol}$ decreases drastically at a certain
radius, except for M51.
For M51, the molecular fraction is still high at the maximum radius of our
data, suggesting that the phase transition must occur at a larger radius.
The characteristic behavior of $f_{\rm mol}$ is the same as that found for
edge-on galaxies in Paper I.
For comparison, we show in Fig.2.b the molecular fraction for edge-on galaxies
taken from Paper I.
It is clear that the behavior of $f_{\rm mol}$ is the same in both figures,
regardless of face-on or edge-on.
This implies that the deconvolving method which we used in Paper I is useful.
Fig.2.a and fig.2.b suggest that the phase transition of HI and H$_2$ in
galaxies occurs within a narrow range of radius in any galaxies of Sb and Sc
types.
The molecular front is one of the fundamental features in late type disk
galaxies.

\section{Gas Phase Transition Theory}

The HI-H$_2$ gas phase transition theory in galaxies has been developed by
Elmegreen (1993, hereafter E93), and we apply this theory to the galaxies for
which we have presented the molecular fraction.
We briefly summarize Elmegreen's phase transition theory.

H$_2$ molecules are formed on the surface of dust, while they are destructed by
UV photons.
Crucial parameters for determining $f_{\rm mol}$ are $P$ (interstellar
pressure), $U$ (UV radiation field, noted as $j$ in E93) and $Z$ (metallicity).
Here the amount of dust is assumed to be proportional to the metallicity $Z$,
and the interstellar pressure $P$ determines the density of gas in clouds.
It is clear that higher $Z$ and lower $U$ make $f_{\rm mol}$ higher, and since
higher density enhances the probability of encounter between HI atoms and
dusts, higher $P$ also makes $f_{\rm mol}$ higher.

Consider a spherical cloud of mass $M$.
Clouds are assumed to be divided into two types according to their internal
structure;
diffuse clouds and self-gravitating clouds.
Diffuse clouds are assumed to be ideal gas without any internal structure and
confined by the interstellar pressure.
Self-gravitating clouds are assumed to be isothermal gas of spherical shape and
also assumed to be bounded by interstellar pressure.
The critical mass between two types is defined by $M_{\rm crit}$ which is the
mass at which the average density of diffuse clouds becomes equal to the
average density of self-gravitating clouds.
The critical mass is about $10^4M_\odot$ in the solar neighborhood, and it of
course varies with the interstellar pressure $P$.
The temperature of clouds are assumed to be constant through the galactic disk,
200K for diffuse clouds and 10K for self-gravitating clouds.
These are standard values for the clouds in the Milky Way Galaxy.

Smaller clouds contain no molecules, because UV photons from outside can reach
even to the center of the cloud and destroy all molecules.
On the other hand, larger clouds can contain more amount of molecules in the
inner part where no UV photon can reach.
Thus the threshold mass for clouds which can contain molecules changes with
$P$, $U$ and $Z$.
As for clouds which are large enough to contain molecules, we can divide the
cloud into two parts at a certain radius; the inner molecular core and the
outer atomic layer.
The radius at which the cloud is divided into two parts is determined by $P$,
$U$, and $Z$, as well as by the cloud's mass.
In this way we can calculate the molecular mass of each cloud for a given set
of $P$, $U$, $Z$ and $M$.

To obtain the molecular fraction in galactic scale, we have to take into
account all clouds with various masses.
The mass distribution of clouds can be approximated by a power law and we take
the mass function with the power law index of -1.5 (Combes 1991) with the lower
cutoff mass of $M_{\rm L}=1 M_\odot$ and the upper cutoff mass of $M_{\rm
U}=10^6 M_\odot$, following E93.
The number of clouds in the mass interval ($M$, $M+dM$) is then given by,
\beq
\phi_{\rm c}dM=\phi_{\rm c0} M^{-1.5}dM
\eeq
Here $\phi_{\rm c0}$ is a constant, and it is determined so that the
integration of this mass function makes total gas density as below,
\beq
\rho = \int_{M_{\rm L}}^{M_{\rm U}} \phi_{\rm c0}MM^{-1.5} dM
\eeq
Using this mass function we can calculate $f_{\rm mol}$ in galactic scale by
integrating all clouds as,

\beq
f_{\rm mol}=\frac{\int_{M_{\rm L}}^{M_{\rm U}} \phi_{\rm c0}M^{-1.5} M_{\rm
mol}dM}{\int_{M_{\rm L}}^{M_{\rm U}} \phi_{\rm c0}M^{-1.5} M dM},
\eeq
where $M$ and $M_{\rm mol}$ are the mass of the cloud and the mass of molecular
core of the cloud respectively, and $M_{\rm L}$ and $M_{\rm U}$ are taken to be
$1M_\odot$ and $10^6M_\odot$, as mentioned above.

Since the resulting $f_{\rm mol}$ changes with $P$, $U$, and $Z$, it changes
with the galacto-centric radius $R$, because $P$, $U$, and $Z$ also change with
the radius in real galaxies.

\section{The Molecular Front in a Model Galaxy}

Using the phase transition theory described above, we calculate $f_{\rm mol}$
as a funtion of galacto-centric radius $R$.
To calculate $f_{\rm mol}$, we have to specify $P$, $U$ and $Z$ as a function
of radius.
We assume that $P$, $U$ and $Z$ can be given by exponential functions of a
radius.
Although this assumption seems to be rough, several evidences support it;
the surface brightness of spiral galaxies in the blue band can be well fitted
by an exponential function (e.g. van der Kruit \& Searle 1982; Elmegreen \&
Elmegreen 1984).
Since the distribution of the blue light is thought to come from young stars
and UV photons come mainly from young stars, $U$ can be approximated by an
exponential function of the scale length as that of the blue band intensity
distribution.
The metallicity $Z$ is known to decrease exponentially with a radius from the
observations of many HII regions in our Galaxy and nearby face-on spiral
galaxies (e.g. D\'iaz 1989; Belly \& Roy 1992; Vila-Costas \& Edmunds 1992).
The interstellar pressure $P$ has not been investigated well, but if we assume
the equipartition between the interstellar hot gas and the random motion of
clouds, we can estimate the interstellar pressure as follows,
\beq
P\sim \rho_{\rm gas}\sigma^2,
\eeq
where $\rho_{\rm gas}$ is local gas density and $\sigma$ is the velocity of
random motion of clouds.
Since $\sigma$ is about 10 km/s and does not changes so much in our Galaxy, we
may assume that $\sigma$ is constant in our Galaxy, and is also so for other
galaxies.
Then we can obtain the interstellar pressure $P$ easily as follows,
\beq
P\propto \rho_{\rm gas}
\eeq
According to Young \& Scoville (1982), the H$_2$ gas density for Sc galaxies is
fitted well by an exponential function.
Since the H$_2$ gas is dominant in the inner region for which we want to
explain the behavior of $f_{\rm mol}$,
the total gas density is roughly approximated by an exponential function.
Then, we may assume that the interstellar pressure $P$ also obeys an
exponential law.

Following Elmegreen(1993), we scale $P$,$V$ and $Z$ with the values of solar
neighborhood; $P_\odot$, $U_\odot$, and $Z_\odot$, and express the three
parameters as a function of galacto-centric radius as follows,
\beq
(P/P_\odot)=\frac{P_0}{P_\odot}\exp(-R/R_P)
\eeq
\beq
(U/U_\odot)=\frac{U_0}{U_\odot}\exp(-R/R_U)
\eeq
\beq
(Z/Z_\odot)=\frac{Z_0}{Z_\odot}\exp(-R/R_Z)
\eeq
where $R_P$, $R_U$ and $R_Z$ are effective radii of each parameter, and
subscript 0 means the value at the galactic center.
It is difficult to determine $P_0$, $U_0$, and $Z_0$ directly, so we define
$R_{P \rm s}$, $R_{U \rm s}$ and $R_{Z \rm s}$, at which $(P/P_\odot)$,
$(U/U_\odot)$ or $(Z/Z_\odot)$ equals to unity.

Substituting $P(R_{P \rm s})=P_\odot$, $U(R_{U \rm s})=U_\odot$, and $Z(R_{Z
\rm s})=Z_\odot$ into above equations, $P$, $U$ and $Z$ are written as follows;

\beq
(P/P_\odot)=\exp(-(R-R_{P \rm s})/R_P),
\eeq
\beq
(U/U_\odot)=\exp(-(R-R_{U \rm s})/R_U),
\eeq
\beq
(Z/Z_\odot)=\exp(-(R-R_{Z \rm s})/R_Z).
\eeq
So, what we have to know are six parameters, $R_P$ and $R_{P \rm s}$ for
calculating $P$, $R_U$ and $R_{U \rm s}$ for $U$, and $R_Z$ and $R_{Z \rm s}$
for $Z$.

In real galaxies, necessary parameters for calculating $f_{\rm mol}$ must be
given independently of each other.
However, they may be related to each other through the interstellar physics and
evolution of the galaxy, and it is likely that the scale length of each
parameters are similar;
for example, the optical scale length and the scale length of molecular gas
density are similar to each other (Young \& Scoville 1982), and the optical
scale length and the scale length of metallicity gradient are similar (D\'iaz
1989; Zaritsky et al. 1994).
In order to know the behavior of $f_{\rm mol}$ in a general case rather than a
certain galaxy, we first consider the simplest case;
 we assume that $R_P=R_U=R_Z$ and $R_{P \rm s}=R_{U \rm s}=R_{Z \rm s}$, and
define $R_{\rm eff}$ and $R_{\rm s}$ as follows;
\beq
R_{\rm eff}\equiv R_P=R_U=R_Z,
\eeq
\beq
R_{\rm s}\equiv R_{P \rm s}=R_{U \rm s}=R_{Z \rm s}.
\eeq
The former equation means that the scale radii of $P$, $U$ and $Z$ are
identical to each other, and the latter means that $f_{\rm mol}=f_{\rm mol
\odot}$ at $R=R_{\rm s}$.

\cen{Fig. 3.a - 3.b}

Under these assumptions, we have calculated the variation of $f_{\rm mol}$ in a
model galaxy which has an exponential disk.
We show the result in Fig.3.a, where we have plotted four lines which
correspond to the case of $R_{\rm s}/R_{\rm eff}=$1, 2, 3 and 4.
In all cases, the observed characteristics of $f_{\rm mol}$ are reproduced
well;
$f_{\rm mol}$ at the galactic center is almost unity,
$f_{\rm mol}$ at the outer region is almost 0,
 and the transition between HI and H$_2$ occurs rather abruptly within a narrow
range of radius, which we have called the molecular front.
It is remarkable that the smooth variations of $P$, $U$ and $Z$, which are
assumed to obey an exponential function with the same scale length, indeed
reproduce such a drastic change of $f_{\rm mol}$.
We note that the molecular front always appears regardless of the ratio of
$R_{\rm s}$ to $R_{\rm eff}$.
This result implies that the molecular front may be one of the fundamental
features of galaxies which have exponential disks.

To investigate how each parameter $P$, $U$, and $Z$ contributes to $f_{\rm
mol}$, we consider the extreme cases in which one of the three parameters ($P$,
$U$, $Z$) is assumed to be constant over the entire disk (equal to the value in
the vicinity of the Sun).
We show the result in Fig.3.b for the case of $R_{\rm s}/R_{\rm eff}=3.5$ but
one of ($P$, $U$, $Z$) is fixed to be equal to the solar value.
It is clear from Fig.3.b that $f_{\rm mol}$ changes its shape if one of ($P$,
$U$, $Z$) is assumed to be constant, and that all of the three parameters
contribute to $f_{\rm mol}$ to some extent.
However, the most striking feature in Fig.3.b is the case in which $Z$ is
assumed to be constant.
In such a case $f_{\rm mol}$ does not reach unity even at the galactic center
and the slope of $f_{\rm mol}$ is extremely flat through the entire disk,
showing no front-like features at all.
If we assume that the metallicity $Z$ is constant of 2$Z_\odot$, then $f_{\rm
mol}$ increases systematically at any radius while keeping the slope nearly
identical.
On the other hand, there still remains front-like feature in the case of
constant $P$ or $U$.
This result implies that $f_{\rm mol}$ depends much more on $Z$ than $P$ or
$U$, and that the metallicity gradient is crucial to the formation of the
molecular front.
Since almost all disk galaxies are known to have metallicity gradient (Belly \&
Roy 1992; Vila-Costas \& Edmunds 1992; Zaritsky et al. 1994), we may say that
the molecular front is one of the characteristic features resulting from the
chemical evolution of disk galaxies.
We note that the strong dependence of $f_{\rm mol}$ on $Z$ is due to the facts
that the metal absorbs the UV photons and shields H$_2$ molecules from UV
destruction as well as it works as a catalysis for the formation of H$_2$
molecules.

\section{ Application to Real Galaxies }
\subsection {Face-on Galaxies}

Now we try to reproduce the variation of $f_{\rm mol}$ in real galaxies.
We give the distribution of $P$, $U$, and $Z$ independently and see whether the
model with a more realistic distribution of $P$, $U$ and $Z$ affects the
behavior of $f_{\rm mol}$.
This will be done more easily for face-on galaxies because the metallicity
distributions, which are crucial to reproduce $f_{\rm mol}$, are available.
So, we try first to reproduce $f_{\rm mol}$ in face-on galaxies for which we
have presented the gas distribution and the molecular fraction in section 2.

Since we have the data of gas density distribution, we can better estimate the
interstellar pressure $P$ using the relation $(P/P_\odot)\sim(\rho_{\rm
gas}/\rho_{\rm gas \odot})$ rather than using an approximated distribution in
the form of an exponential function.
Furthermore, the surface brightness distribution can be obtained from optical
observation, and the metallicity and its gradient also can be obtained from
optical spectroscopy.
Thus, we can give all parameters independently and investigate whether the
model can describe the actual behavior of $f_{\rm mol}$ in galaxies.
The parameters used here and the references from which the data have been taken
are summarized in Table 2.

\cen{Fig. 4.a - 4.d}

We show the results from the model calculation in Fig.4.a-4.d, superposed on
the observed $f_{\rm mol}$.
As seen in Fig.4, $f_{\rm mol}$ predicted by the model can reproduce the
observed one.
In the case for M101, NGC6946 and IC342, the model predict the molecular front
clearly, and its shape and location agrees well with the observation although
there exists slight deviation.
In the case for M51, although the observed $f_{\rm mol}$ is almost unity within
the region where the data are available and show no front-like feature, the
model reproduce the observed $f_{\rm mol}$ well again.
The model predicts rather flat $f_{\rm mol}$, almost constant unity in this
region and no molecular front at least in this region.
This result comes from the fact that the metallicity in this region is very
high (about 3.5$Z_\odot$ according to Vila-Costas \& Edmunds 1992) and constant
up to 6 kpc.
As described in section 4, the metallicity gradient is crucial to the existence
of the molecular front, and we have shown that the molecular front would not
appear in a galaxy without metallicity gradient (Fig.3.b).
The case of M51 just corresponds to this case but higher metallicity.
It is known that the metallicity decreases exponentially in the outer region of
M51 where we do no concern here (Vila-Costas \& Edmunds 1992).
So this galaxy would have the molecular front at large radius.
(see Kuno et al 1995 for a more detailed study of $f_{\rm mol}$ in M51)

Although the model reproduces the qualitative behavior of $f_{\rm mol}$, there
still exists residual to be explained.
We think it comes from both the observational error and the model.
As mentioned above, we give $U$ and $Z$ by exponential functions, but it is
likely that the true distribution of $U$ or $Z$ deviates from an exponential
function (locally and/or globally.)
For instance, the metallicity gradient in the inner region of M51 and M101
differs from that in the outer region (Vila-Costas and Edmunds 1992).
This result implies that the metallicity gradient in a form of exponential
function is not universal, but sometimes not appropriate.
So a more precise observation of the distribution of the metallicity is
necessary for further argument.
In addition to this, the quantitative disagreement of $f_{\rm mol}$ also comes
from the use of the optical scale length at the blue band instead of the U
band.
In blue band, the contribution of evolved giants to the luminosity profile is
not entirely negligible, and the surface brightness may not trace correctly the
distribution of young OB starts, while this is not the case in U-band in which
young OB stars dominate the luminosity.
So we need the data of U band surface brightness distribution with high quality
for further improvement.

On the other hand, we also think that the model itself contains some
uncertainties.
The phase transition theory by Elmegreen (1993) includes several parameters
which have not been determined precisely yet; for example, the power low index
of the clouds' mass function, lower and upper limit masses for the clouds' mass
function, the temperature of clouds, and so on.
Their values used in the theory are thought to be common now, but they might be
changed by further observation.

\subsection{Edge-on Galaxies}

For edge-on galaxies, for which $f_{\rm mol}$ have been obtained in Paper I,
there are some difficulties in obtaining the accurate values of necessary
parameters, especially parameters which are related to the metallicity (the
metallicity and its gradient) because they have been measured only in nearby
face-on spiral galaxies.
Here we estimate the metallicity by using empirical relations, since the
metallicity gradient normalized by the optical scale length of the disk is
almost constant through galaxies (D\'iaz 1992; Zaristky et al. 1994), and the
characteristic abundance of galaxies log(O/H)$_{0.8 R_B}$, which is defined as
the metallicity at the radius of 0.8$\times R_B$ ($R_B$; the optical scale
length at the B band) is correlated to the mass of galaxies (Zaritsky et al.
1994).
These empirical relations are shown in Fig.5.a and Fig.5.b, in which the data
are taken from Zaritsky et al. (1994).
\cen{Fig.5.a - 5.b}
We have obtained the empirical relation between the characteristic abundance
and the rotation velocity of galaxies, and the relation between the metallicity
gradient and the rotation velocity by applying a least square fit.
The resulting relations are expressed as follows;
\beq
{\rm [12 + log(O/H)]}_{0.8 R_B}=8.53 \pm 0.17 + 0.0244 V_{\rm rot},
\eeq
\beq
{\rm \Delta[log(O/H)]}/R_B =  -0.196 \pm 0.105 + 6.81\times 10^{-5}  V_{\rm
rot}.
\eeq

The first relation shows that more massive galaxy has larger characteristic
abundance.
This tendency is consistent with the results obtained by Roberts \& Haynes
(1994).
In the second relation, the dependence of the gradient on the rotation velocity
is small and the gradient is almost constant.
The constant value of -0.196 is consistent with the result by D\`iaz (1989)
when we consider the scatter.
Since these relations are not unique and have some scatter, we cannot give a
unique value of $R_Z$ or $R_{Z \rm s}$ from these empirical relation.
We fix the metallicity gradient to be equal to the moderate value in the second
relation, while two possible characteristic abundances are considered;
the maximum and the minimum characteristic abundance which correspond to
$\pm$1$\sigma$ in the first relation.
So the metallicity gradients in two cases are the same but the total amount of
metal is different from each other.

The data of surface brightness distribution for NGC 3079 have been taken from
Shaw et al. (1993), and the data for other galaxies from van der Kruit \&
Searle (1982).
The interstellar pressure $P$ is estimated from the gas distribution, in the
same way done in the former subsection.
The parameters for these galaxies are summarized in Table 3.

\cen{Fig. 6.a - 6.c}

We show thus obtained $f_{\rm mol}$ for NGC 891, NGC 3079, and NGC 4565 in
Fig.6.a-6.c, superposed on the observed $f_{\rm mol}$ which have been obtained
in Paper I.
We exclude NGC 5907 because the HI data for this galaxy has not been obtained
over the hole disk and it might cause for substaintail error in estimating the
interstellar pressure $P$, especially in the outer galaxy where HI gas is
dominant.
In Fig.6.a, we show the results for NGC 891.
The model reproduces the variation of $f_{\rm mol}$ well again, as seen for the
face-on galaxies.
The observed $f_{\rm mol}$ lies between two plots and it is slightly nearer to
the case of the higher characteristic abundance.
So the characteristic abundance in this galaxy may be really higher than the
moderate value in the relation obtained above.
For the case of NGC 3079, the observed $f_{\rm mol}$ is also in the possible
range predicted by the model and the model can reproduce $f_{\rm mol}$ well.
The scatter of the empirical relation causes less difference of $f_{\rm mol}$
than that for NGC 891, because the molecular gas is highly concentrated in the
central region of this galaxy (Paper I; Sofue \& Irwin 1992) and the effect of
the interstellar pressure $P$ must be larger than the effect of the scatter of
the  $Z$.
It is remarkable that the present model is applicable to such a peculiar galaxy
with nuclear activity.
This implies that the interstellar physics in the disk of active galaxy such as
NGC 3079 must be the same as the normal disk galaxies.
In Fig.6.c we show the models for NGC 4565.
The model can explain the qualitative characteristic of $f_{\rm mol}$ for NGC
4565, including the sudden decrease at R$\sim$3 kpc and the enhancement at R
$\sim$ 5kpc.
The $f_{\rm mol}$ calculated by the model for NGC 4565 is smaller than the
observed $f_{\rm mol}$ throughout the disk, but if we give a slightly steeper
gradient, it is possible to reproduce the observed $f_{\rm mol}$, because
$f_{\rm mol}$ is so sensitive to $Z$, and its value may lie in the possible
range in Fig.5.b.

The observed $f_{\rm mol}$ for edge-on galaxies has been reproduced well using
moderate metallicity gradient which is determined by the empirical relation.
In the former subsection we have already seen that the observed $f_{\rm mol}$
for face-on galaxies has also been explained well using the set of ($P$, $U$,
$Z$) determined observationally.
 From these results we stress that the model described in the present paper
works well to explain the phase transition of HI and H$_2$ gases in galaxies.

\section{On the Relation to the Chemical Evolution of the Galaxy}

As described in Section 3.2, $f_{\rm mol}$ depends strongly on the metallicity
Z and the metallicity gradient is crucial to the formation of the molecular
front.
These results imply that the gas distribution observed at present is tightly
related to the history of the chemical evolution of a galaxy.
Since the disk of spiral galaxy was virtually metal-free at the earliest stage
of the evolution and the metallicity has been increasing monotonically with
time, $f_{\rm mol}$ at any place in the disk must have changed with time.
Then, the question is how $f_{\rm mol}$ has varied with time in a galaxy, and
when and how the molecular front appeared and evolved since then.
Since there was no metallicity gradient in the disk at the early stage, the
molecular front did not exist at that time.
It must have formed with the growth of the metallicity gradient, but when and
how it has been formed is still a matter of debate because it is not known yet
how the metallicity gradient formed in a disk galaxy.

Once the molecular front is formed, it would not disappear unless the
metallicity gradient would vanish.
However, the molecular front would not stay at the same position, instead it
should  change its shape and location as the chemical evolution proceeds in a
galaxy.
For example, suppose that the metallicity increases in global while keeping the
slope of the gradient constant.
In this case, the molecular front advances outwards with time without changing
its shape; this is seen in Fig.3.a.
The most chemically-evolved case corresponds to the line with $R_{\rm s}/R_{\rm
eff}$=4, while the least enriched case corresponds to the case of $R_{\rm
s}/R_{\rm eff}=1$.
In real galaxies the slope of the metallicity gradient may also change.
However, there is no doubt that the metallicity has increased monotonically, so
we think the molecular front may advance outwards with time.

Though it is still controversial whether the star formation rate is more
tightly related to HI gas or to H$_2$ gas, it is interesting to consider the
effect of $f_{\rm mol}$ on the evolution of galaxies through star formation.
We here assume that the star formation rate is determined by the amount of
H$_2$ gas, not by HI gas.
More precisely, we should assume that the star formation rate is determined by
the amount of molecular gravitating clouds rather than total amount of
molecular clouds which include both gravitating and diffuse molecular clouds,
because star formation would rarely occur in diffuse clouds.
However, as far as our model is concerned, more than 80\% of H$_2$ gas is made
up from self-gravitating clouds in any case, so it is a good approximation to
assume that star formation rate is determined by the amount of H$_2$.
If we assume a Schmidt's law, $SFR\propto n({\rm H_2})^{\alpha}$, then $f_{\rm
mol}$ represents somehow the star formation efficiency from total gas.
So the variation of $f_{\rm mol}$ through the disk manifests the star formation
efficiency in the disk;
star formation efficiency is high in the inner disk and many stars are forming
or will form soon there, while few stars are forming in the outer region.
So, the molecular front divides the disk into two parts; the inner star forming
disk and the outer disk where stars are rarely formed.
This is indeed observed in NGC 891 (see Sofue \& Nakai 1993).

Once stars are formed under a condition of certain $f_{\rm mol}$, newly formed
OB stars radiate UV photons and soon explode as supernovae, ejecting heavy
elements.
So some 10$^{6\sim7}$ years after the star formation, the physical condition of
ISM must be changed by the newly formed stars.
This implies the change of ($P$,$U$,$Z$), and so the change of $f_{\rm mol}$
and star formation efficiency.
A galaxy must evolve with such a cycle; a given set of ($P$,$U$,$Z$) determine
$f_{\rm mol}$  and so the star formation efficiency, and then newly formed
stars under such condition of star formation changes the physical condition;
$P$, $U$ and $Z$, leading to different $f_{\rm mol}$ and star formation
efficiency.

 From such a view of galaxy evolution, it is interesting to consider the star
formation history at the earliest stage of evolution of a galaxy.
The first generation of stars must have been formed from the unenriched
molecular cloud in the central region of a galaxy, where H$_2$ molecules had
been formed by a direct triple encounter of H atoms.
(We do not take into account this process for calculation of $f_{\rm mol}$
because the formation of H$_2$ by dust is much more efficient than direct
encounter, but in case of no metal, H$_2$ molecules must have been formed by
such process.)
Once stars formed, they emit UV radiation and destruct molecules quite
efficiently compared with the molecular clouds at the present epoch because of
the lack of absorbing dust.
So star formation efficiency soon decreases with the star formation, and so
only few stars are formed after the first star formation.
Then some 10$^{6\sim 7}$ years later all of OB stars of the first generation
explode as supernovae, ejecting heavy elements, and this makes $U$ lower and
$Z$ higher, and so $f_{\rm mol}$ higher, leading to the formation of second
generation of stars.
A galaxy at very early stage must have evolved with such a drastic change of
$f_{\rm mol}$ until the gas has been enriched enough to protect clouds from the
UV radiation.

In summary, in order to investigate the evolution of galaxy more precisely, the
atomic-to-molecular phase transition process is important through the star
formation.
So, it would be of vital importance to construct models which include such a
process instead of the simple star formation low such a Schmidt law, in which
the star formation rate is simply assumed to proportional to the total amount
of gas.

\vspace{2cm}
\section*{Acknowledgement}

This work was financially supported in part by a Grant-in-Aid for the
Scientific Research No.06233206 and No.06640349 by the Japanese Ministry of
Education, Culture and Science.

\newpage
\section*{References}

\def\r{\hangindent=1pc  \noindent}

\r Belly, J., Roy. J.-R. 1992, ApJS, 78, 61.

\r Combes, F., 1991, ARA\&A, 29, 195.

\r D\'iaz, A. I. 1989, in Evolutionary Phenomena in Galaxies, eds. J. E.
Beckman, B. E. J. Pagel (Cambridge Univ. Press, Cambridge), p. 377.


\r Elmegreen, B. G. 1993, ApJ 411, 170.

\r Elmegreen, D. M., Elmegreen, B. G., 1984, ApJS, 54, 127.


\r Hollenbach, D., Werner, M., Salpeter, E., 1971, ApJ, 163, 165.


\r Irwin, J. A., Seaquist, E. R., 1991, ApJ, 371, 111.

\r Kenney, J. D. P., Scoville, N. Z., Wilson, C. D., 1991, ApJ, 366, 432.

\r Kuno, N., Nakai, N., Handa, T., Sofue, Y., 1995, PASJ, to be submitted.

\r Nakai, N., Kuno, N., Handa, T., Sofue, Y., 1994, PASJ, 46, 527.

\r Roberts, M. S., Haynes, M. P., 1994, ARA\&A, 32, 115.

\r Rots, A. H., Bosma, A., van der Hulst, J. M., Athanassoula, E., Crane, P.
C., 1990, AJ, 100, 387.

\r Rupen, M. P., 1991, AJ, 102, 48.

\r Sandage, A., Tammann, G. A., 1974, ApJ, 194, 559.


\r Shaw, M., Wilkinson, A., Carter, D., 1993, A\&A, 268, 511

\r Sofue, Y., 1994, PASJ 46, 173. 

\r Sofue, Y., Honma, M., Arimoto, N., 1995, A\&A, in press.

\r Sofue, Y., Irwin, J., A., 1992, PASJ, 44, 353.

\r Sofue, Y., Nakai, N., 1993, PASJ 45, 139. 

\r Sofue, Y., Nakai, N., 1994, PASJ 46, 147. 


\r van der Kruit, P. C., Searle, L., 1982, A\&A, 110, 61.

\r Vila-Costas, M. B., Edmunds, M. G., 1992, MNRAS, 259, 121.

\r Young, J. S., Scoville, N., 1982, ApJ, 258, 467.

\r Zaritsky, D., Kennicutt, R. C. Jr., Huchra., J. P., 1994, ApJ, 420, 87.

\newpage
\section*{Figure captions}

{\bf Fig.1.a} The surface mass distribution of HI and H$_2$ gases in M51.
The amount of H$_2$ gas is calculated from the intensity of CO line, assuming
the constant conversion factor of 3$\times$10$^{20}$.
The references from which the data have been taken are listed in Table 1.

{\bf Fig.1.b} Same as Fig.1.a, but for M101.

{\bf Fig.1.c} Same as Fig.1.a, but for NGC 6946.

{\bf Fig.1.d} Same as Fig.1.a, but for IC 342.

{\bf Fig.2.a} The molecular fraction $f_{\rm mol}$ in face-on galaxies.

{\bf Fig.2.b} The molecular fraction $f_{\rm mol}$ for edge-on galaxies, which
has been obtained in Paper I.

{\bf Fig.3.a} The molecular fraction $f_{\rm mol}$ for a model galaxies in
which all of $P$, $U$, and $Z$ obey an exponential function with same scale
length of $R_{\rm eff}$. These four lines correspond to the case that the ratio
of $R_{\rm s}/R_{\rm eff}$ equals to 1, 2, 3 and 4.
In all cases $f_{\rm mol}$ changes rather drastically at a certain region,
which we call molecular front.

{\bf Fig.3.b} Models without gradient of one of three parameters $P$, $U$, and
$Z$.
The solid line corresponds to the model same as in Fig.1 with $R_{\rm}/R_{\rm
eff}=3.5$, and other three lines correspond to the cases of constant $P$, $U$,
or $Z$.
It is remarkable that the metallicity $Z$ is the most dominant parameter.

{\bf Fig.4.a} $f_{\rm mol}$ for M51 by the model, superposed on the observed
$f_{\rm mol}$.
The necessary parameters are given observationally and are listed in Table 2.
The vertical bar indicates typical observational error of $f_{\rm mol}$, and
the error in the inner region is less than that.

{\bf Fig.4.b} same as Fig.4.a, but for M101.

{\bf Fig.4.c} same as Fig.4.a, but for NGC 6946.

{\bf Fig.4.d} same as Fig.4.a, but for IC 342

{\bf Fig.5.a} The empirical relation between the characteristic abundance and
the rotation velocity of galaxies.
Characteristic abundance is defined as the abundance at R=0.8$R_B$ where R$_B$
is the scale length of the optical disk.
The solid line shows the empirical relation given by least-square fit, and two
dashed lines show the scatter of 1$\sigma$.

{\bf Fig.5.b} The empirical relation between the metallicity gradient
normalized by the scale length of the optical disk and the rotation velocity.
Note that the gradient is almost constant through galaxies.
The notation of solid and dashed lines are same as Fig.5.a.

{\bf Fig.6.a} $f_{\rm mol}$ calculated for NGC 891, superposed on observed
$f_{\rm mol}$ from Paper I.
The dotted line corresponds to the case with the moderate $R_Z$ and $R_{Z \rm
s}$, and the heavy solid line corresponds to the case of the maximum $R_Z$ and
$R_{Z \rm s}$ (1$\sigma$).
The observed $f_{\rm mol}$ (thin line) lies between two plots, but is much
nearer to the case of of the maximum $R_Z$ and $R_{Z \rm s}$

{\bf Fig.6.b} Same as Fig.6.a, but for NGC 3079.

{\bf Fig.6.c} Same as Fig.6.a, but for NGC 4565.

\newpage

Table 1 : Assumed distances and references for four face-on galaxies.
\vs\vs

\setlength {\tabcolsep}{0.3cm}

\begin{tabular}{cccc}
\hline\hline
galaxy  & Distance& references for distance & references for HI and CO \\
\hline
M 51    & 9.6 Mpc & a & c,d \\
M 101   & 4.1 Mpc & b & e \\
NGC 6946& 5.9 Mpc & b & f \\
IC 342  & 2.7 Mpc & b & f \\
\hline\\
\end{tabular}
\vs\vs

a. Sandage \& Tammann (1974).
\vs

b. Vila-Costas \& Edmunds (1992).
\vs

c. Rots et al. (1990).
\vs

d. Nakai et al. (1994).
\vs

e. Kenney et al. (1991).
\vs

f. Young \& Scoville (1982)

\vs\vs

\newpage

\setlength {\tabcolsep}{0.9cm}

Table 2 : The parameters for four face-on galaxies.
\vs\vs

\begin{tabular}{ccccc}
\hline\hline
galaxy & $R_U$ $^{\rm a}$ & $R_{U \rm s}$ $^{\rm a}$ & $R_Z$ & $R_{Z \rm s}$ \\
\hline
M 51     & 5.2  & 11.5  & --- & --- $^{\rm b}$ \\
M 101    & 5.2  & 12.3  & 3.3 & 5.2 $^{\rm c}$ \\
NGC 6946 & 4.0  & 6.7   & 4.9 & 6.5 $^{\rm d}$ \\
IC 342   & 5.5  & 2.9   & 5.0 & 5.9 $^{\rm c}$ \\
\hline
\end{tabular}
\vs\vs

a. Elmegreen \& Elmegreen (1984).
\vs

b. The metallicity in the inner region of M51 is constant, equal to
3.5$Z_\odot$ (Vila-Costas \& Edmunds 1992).
\vs

c. Vila-Costas \& Edmunds (1992).
\vs

d. Belly \& Roy (1992).

\vs\vs
\newpage

\setlength {\tabcolsep}{0.9cm}

Table 3 : The parameters for three edge-on galaxy.
\vs\vs

\begin{tabular}{cccc}
\hline\hline
galaxy & Distance $^{\rm a}$ & $R_U$ & $R_{U \rm s}$ \\
\hline
NGC 891   &  8.9 Mpc  & 4.6 kpc  & 5.0 kpc $^{\rm b}$ \\
NGC 3079  & 15.0 Mpc  & 4.7 kpc  & 7.3 kpc $^{\rm c}$\\
NGC 4565  & 10.2 Mpc  & 5.5 kpc  & 9.5 kpc $^{\rm b}$\\
\hline\\
\end{tabular}
\vs\vs

a. Paper I.
\vs

b. van der Kruit \& Searle (1982).
\vs

c. Shaw et al. (1993).
\vs

\end{document}